# Bistable electric field control of single-atom magnetocrystalline anisotropy

**Authors**

Jose Martinez-Castro[1,2,3*], Cyrus F. Hirjibehedin[1,2,4†], and David Serrate[3,5*]

**Affiliations**

[1] London Centre for Nanotechnology, University College London (UCL), London WC1H 0AH, UK.

[2] Department of Physics & Astronomy, UCL, London WC1E 6BT, UK.

[3] Instituto de Nanociencia de Aragón and Laboratorio de Microscopías Avanzadas, Universidad de Zaragoza, 50018 Zaragoza, Spain.

[4] Department of Chemistry, UCL, London WC1H 0AJ, UK.

[5] Departamento de Física de la Materia Condensada, Universidad de Zaragoza, 50009 Zaragoza, Spain.

[*]Correspondence to: serrate@unizar.es, j.martinez@fz-juelich.de
[†]present address: MIT Lincoln Laboratory, Lexington, MA 02421-6426, USA

**Abstract**

We reversibly switch the polar environment of an individual magnetic atom with an electric field to control the energy barrier for reversal of magnetization. By applying an electric field in the gap between the tip and sample of a scanning tunneling microscope, we induce bistable changes in the polarization of the region surrounding a chlorine vacancy in a monolayer of sodium chloride on copper terminated by a monolayer of copper nitride. The displacement of the sodium chloride ions alters the local electric polarization and modifies the magnetocrystalline anisotropy experienced by a single cobalt atom. When a cobalt atom is near a chlorine vacancy, spin-sensitive inelastic electron tunneling spectroscopy measurements can reveal the change in anisotropy. The demonstration of atomic-scale control of magnetic properties with electric fields opens new possibilities for probing the origins of magnetoelectric coupling and will stimulate the development of model artificial mutliferroic systems.

**MAIN TEXT**

**Introduction**

Achieving electric field control of magnetic properties is a major challenge in the development of novel materials and devices, offering access to new material properties as well as potential technological improvements such as significantly reduced power consumption (*1*). A variety of driving mechanisms to couple a material's electronic and magnetic degrees of freedom have been proposed. For example, an electrostatic gate voltage can force electronic transport in quantum systems to proceed through discrete spin states with a well-defined conductivity (*2–5*). In thin ferromagnetic metals and semiconductors, the charge redistribution near a strong gating electric field can significantly alter the magnetic coercivity (*6*) or ordering temperatures (*7*). In addition, spin-orbit coupling enables magnetization control by electrical currents through spin-torque effects (*8–10*). Alternatively, direct coupling of electrostatic and magnetic degrees of freedom has been achieved in single phase multiferroic materials like $BiFeO_3$ (*11*) and hexagonal manganites (*12*,

*13*), or in heterostructures interfacing ferroelectric and magnetic thin films (*14*, *15*). In spite of the intense research activity on multiferroic phenomena, a framework enabling fundamental studies on the coupling of the electric polarization and the spin moment at interfaces is not yet fully developed (*15*).

Magnetocrystalline anisotropy energy (MAE) is one of the most relevant parameters for defining the behavior of magnetic materials. It determines the susceptibility of a material's magnetization to thermal activation, external magnetic fields, and electromagnetic radiation (*16*). Different studies have demonstrated that MAE in metallic thin films can be continuously tuned at the nanoscale by the application of an electric field (*6*, *17*, *18*). In these cases, the coupling mechanism is restricted to the response of the electronic density of states to the unscreened part of the electric field in the metal. At the nanoscale, the electric field can influence the MAE barrier and modify the magnetization reversal attempt frequency in the superparamagnetic regime (*18*).

In the fundamental limit of an individual magnetic atom, the atom's MAE is mainly controlled by the structure and charge distribution of the immediate environment (*19–21*). At the single molecule level, it is possible to modify the MAE by charging (*22*) as well as through controlled (reversible or non-reversible) deformation of the bond distance between the ion carrying the magnetic moment and its surrounding ligands (*23–25*). Analogously, an observable change in MAE can be achieved by local strain arising from deformation of the substrate supporting the magnetic moments (*26*). This suggests that controlling the arrangement of the atoms surrounding a magnetic ion using an electric field would also modify the single-atom MAE, resulting in an efficient way to implement external electric field control of magnetism.

In this work, we show that the MAE of an individual magnetic atom can be manipulated through bistable atomic displacements controlled by an external electric field applied to a supporting dipolar substrate. By depositing a monolayer (ML) of NaCl on the atomically thin polar insulator copper nitride ($Cu_2N$) capping bulk Cu(001), we induce a distortion in the NaCl that results in a net out of plane dipole similar to what has been observed for a bilayer of NaCl on $Cu_2N$ (*27*). In the presence of a Cl vacancy, the NaCl can be bistably switched between two dipolar orientations using an electric field applied from the tip of a scanning tunneling microscope (STM) used to study the system. By performing spin-sensitive inelastic electron tunneling spectroscopy (IETS) on a Co atom adsorbed on the NaCl-ML, we observe that the characteristic MAE measured for Co on bare $Cu_2N$ (*28*, *29*) is altered between two distinct values following the bistable electric polarization of the NaCl-ML that can be programmed by opposite electric fields in the tip-sample gap. These results show that electric field control of magnetic properties can be achieved in the limit of single atoms on surfaces. Extending this technique to other materials in which more detailed characterization can be performed would enable the development of model systems for understanding the interplay between MAE and polar order, shedding light on the atomic-scale origins of multiferroic coupling.

**Bistable polarization switching in a monolayer.** Figure 1A shows a topographic STM image of a NaCl ML covering Cu$_2$N nanoislands on Cu(001). As is seen for ultra-thin films of NaCl on many other substrates (*30*), the NaCl ML on Cu$_2$N also contain Cl vacancies (Fig. 1A). Co atoms can also be deposited on top of the NaCl ML, and as seen in Fig. 1B can be recognized as a bright protrusions similar to those observed for Co adsorbed on top of Cu$_2$N/Cu(001) (*28*). However, the appearance of Co adatoms on the NaCl ML strongly depends on the adsorption site. As seen in Fig. 1C, multiple adsorption sites can be identified for Co, for example on top of Na or Cl sites (Fig S1A and S1B, respectively). In addition, the Co atom has an unusual appearance near a Cl vacancy (Fig. S1), where it does not have the round characteristic shape of a Co atom adsorbed on Cu$_2$N (Fig. 1C) or the four-fold symmetry expected above Cl sites. This irregular shape is attributed to a Co atom because the appearance on high symmetry sites can be systematically and repeatedly recovered by displacing a Co atom using lateral manipulation near a Cl vacancy on the NaCl ML (Fig. S1). Furthermore, its identity can be confirmed because its spectroscopic signature is similar to that of Co on Cu$_2$N (Fig. 3).

As has been observed for the NaCl BL on Cu$_2$N (*27*), the polarization of the NaCl ML can be bistably switched in the presence of a Cl vacancy. When the tip is positioned above a Cl vacancy, the strongly self-poled polarization of the NaCl layer, which is induced by the polar orientation of the underlying Cu$_2$N, can be switched by ramping the bias to positive values, thus applying a positive electric field (i.e. pointing from the sample to the tip) (Fig. 2A). The polarization also can be reversibly switched back to the original state by applying large enough negative electric fields. The sharp change in the tunneling current is attributed to tunneling electro resistance (TER), where the reversal of the dipole orientation modifies the work function of the substrate and therefore the height of the tunneling barrier (*27, 31*).

Simultaneously acquired Kelvin probe measurements obtained using atomic force microscopy (AFM) further confirm the change in the work function of the substrate from dipolar reversal. As see in Fig. 2B, the shift of the resonance frequency as a function of voltage $\Delta f(V)$ shows the expected parabolic behavior (*32*), with the contact potential difference $V_{cpd}$ between the tip and substrate work functions marked by the parabola's maximum. The value of $V_{cpd}$ clearly shifts for the two states, indicating the difference in work function and electric polarization.

**Bistable switching of magnetic anisotropy.** Having seen that switching the Cl vacancy in the NaCl ML results in a change of its electric polarization, we explore the impact of this change on the magnetic properties of Co atoms adsorbed nearby. A Co atom on bare Cu$_2$N has a quantum spin $S=3/2$ (*28*). Because of the anisotropic arrangement of charge in the Cu$_2$N below the magnetic atom (*21*), the crystal field splits the states of spin projection along the z-axis $S_z=1/2$ and $S_z=3/2$ by an energy $E_{an}$ (*19*). These states further split in an applied magnetic field according to the spin Hamiltonian

$$\hat{H} = -g\mu_B \hat{\boldsymbol{B}} \cdot \hat{\boldsymbol{S}} + DS_z^2$$

where $g$ is the Landé factor, $\mu_B$ is the Bohr magneton, and $D$ the is the uniaxial anisotropy parameter. For Co on Cu$_2$N, $D > 0$ so $S_z=\pm 1/2$ is the doubly degenerate ground state. Exchange coupling to the underlying conduction electrons results in Kondo screening of this state (*28*), manifesting as a sharp resonance at the Fermi energy $E_F$ ($V=0$). STM-based IETS induces transitions between the $S_z$ states (*21*), resulting in conductance (d$I$/d$V$) steps when the sample bias matches $E_{an}/e$ at ~ $\pm 5$ mV.

As seen in Fig. 3, a similar spectrum is observed for a Co atom near a Cl vacancy in the NaCl ML: the relative amplitude of the Kondo resonance is reduced and no additional conductance steps (i.e. inelastic spin excitations) are observed. This suggests that the additional NaCl ML above the $Cu_2N$ decreases the exchange coupling between the magnetic impurity and the nearby conducting electrode without changing the total spin of the Co atom (*28*, *29*). Spectroscopic measurements performed on Co atoms adsorbed on Na or Cl sites on top of the NaCl ML but away from Cl vacancies did not show any characteristic IETS features in this energy range. This may be because the inelastic component of the tunneling current, which induces spin-flip excitations, is too small to be resolved for experimental conditions in which Co atoms remains stable.

To study the impact of the polarization switching on the Co atom, we position the tip above a Co atom located near a Cl vacancy and ramp the applied voltage. Current jumps at critical voltages corresponding to critical electric fields (Fig. 4) confirm that polarization switching occurs even with the Co atom present (Fig. 5A). To ensure that no other process happened while setting the conditions for low bias spectroscopy, variations in $V_{set}$ and $I_{set}$ before and after the polarization switch were continuously monitored.

As seen in Fig. 5B, two different spectroscopic signatures can be distinguished for the two different polarization states (labeled A and B). Note that the electric field conditions during the spectroscopy in both states are identical. As demonstrated by the change in voltage of the IETS step, the MAE of the Co atom is decreased by a factor of two when the underlying NaCl ML is switched from state A to B, while the amplitude of the Kondo resonance is enhanced. The similarity of the spectra (i.e. same number of inelastic excitations and the persistence of a Kondo resonance) implies that the value of $S$ and the sign of $D$ remain the same for both states. This excludes a charging process on the Co atom as the origin of the bistable switching.

Additional confirmation of the change in MAE experienced by the Co atom for the two different surface polarization states is obtained by observing the evolution of the spectra with magnetic field (Fig. 6), which is illustrated in Fig. 5C. For the relatively small magnetic fields (up to $B = 3$ T) accessible in these studies, only a very small change in the energy of the inelastic tunneling step is expected (*28*). The small shift that is observed is consistent with such changes. A much more prominent change, however, can be observed in the splitting of the Kondo resonance, which is much more sensitive to changes in magnetic field (*28*). As seen in Fig. 6, the splitting of the Kondo resonance is different for the two different polarization states.

Figure S2 shows low energy spectroscopy for three other Co atoms near Cl vacancies as a function of the polarization. The general features are the same in all cases, though the MAE either increases or decreases from state A to B depending on the local environment. Variations are also observed for changes in the Kondo resonance, which may be modified due to changes in the strength of the coupling between the Co atom and the underlying metallic substrate. It is difficult to quantify these variations because the adsorption site of the Co atom near the Cl vacancy as well as the underlying lattice structure nearby cannot be resolved; this may be due to the low symmetry around the Cl vacancy and the relatively weak adsorption energy of the Co atom, which precludes high-resolution imaging at small tip-sample distances.

**Discussion**

We have demonstrated that electric field induced modification of the polarization of a substrate can bistably switch the magnetocrystalline anisotropy experienced by a single magnetic atom through the rearrangement of the atomic positions of the neighboring ions. As is the case for nanoscale magnetic data storage (*33*), bistable modulation of an individual atom's MAE could have applicates for classical and quantum information processing, potentially allowing for switching to an easily modifiable state when writing data and then back a more stable state for longer-term storage. Further development of atomic manipulation techniques on this or other switchable substrates would also facilitate the construction of coupled spin systems, enabling construction of the smallest possible multiferroic systems in which a collective electric degree of freedom could be used to control the collective magnetic degree of freedom at the atomic scale. If this can be performed on a surface for which the switched polarization state can be fully characterized, then the system would provide an ideal venue for studying the coupling between collective polar and magnetic order at the level of a single atomic spin. This would represent a model system for understanding the fundamentals of multiferroic behavior.

**Materials and Methods**

**Scanning Tunneling Microscopy and Atomic Force Microscopy.** Scanning tunneling microscopy (STM) experiments were performed using a Specs JT-STM, a commercial adaptation of the design described by Zhang et al. (*34*) as well as an Omicron Nanotechnology LT-STM with a qPlus force sensor (*35*) installed for combined operation of both STM and atomic force microscopy (AFM). The qPlus sensor, with a resonance frequency $f_0 = 23379.5$ Hz and a stiffness $k \sim 1800$ N/m (*35*), was operated in non-contact AFM mode with a phase-locked excitation at a constant oscillation amplitude of $A = 20$ pm. Both systems were operated in ultrahigh vacuum conditions, with typical chamber pressures below $2\times10^{-10}$ mbar, and at base temperatures of 1.1 K and 4.5 K respectively. In the Specs JT-STM, a magnetic field up to 3 T can be applied perpendicular to the sample surface.

The bias voltage $V$ is quoted in sample bias convention. Topographic images were obtained in the constant current imaging mode with $V$ and tunnel current $I$ set to $V_{set}$ and $I_{set}$ respectively. Differential conductance $dI/dV$ measurements are obtained using a lock-in amplifier, with typical modulation voltages of 150 µV at ~840 Hz added to $V$. Spectroscopy is acquired by initially setting $V=V_{set}$ and $I=I_{set}$, disabling the feedback loop to maintain position of the tip, and then sweeping $V$ while recording $I$, $dI/dV$, and/or the shift of the resonance frequency $\Delta f$.

Cu(001) samples (MaTeck single crystal with 99.999% purity) were prepared by repeated cycles of sputtering with Ar and annealing to 500 ˚C. $Cu_2N$ is prepared on top of clean Cu(001) samples by sputtering with $N_2$ and annealing to 350 ˚C (*36*). Deposition of NaCl was performed using a Knudsen effusion cell operated at 490 ˚C and with the $Cu_2N$/Cu(001) substrate at room temperature (*27*).

Topographic images obtained using STM were processed using WSxM (*37*).

**Lateral atomic manipulation.** Co atoms on top of Na sites on NaCl on $Cu_2N$/Cu(001) can be laterally manipulated to the next Na atomic position by approaching the STM tip in closed feedback mode to maintain a constant $I_{set}$ and (simultaneously) decreasing $V$. Typical starting conditions are $V_{set}$=-1.3 V and $I_{set}$=10 pA. $V$ is then decreased until a sharp jump in the tip position is observed, typically when $V < 100$ mV. The tip is then moved across the NaCl ML island while the Co follows the tip. The speed of the tip during lateral manipulation is below 10 nm/s.

**References and Notes:**

1. F. Matsukura, Y. Tokura, H. Ohno, *Nat. Nanotechnol.* **10**, 209–220 (2015).
2. J. Paaske et al., *Nat. Phys.* **2**, 460–464 (2006).
3. N. Roch, S. Florens, V. Bouchiat, W. Wernsdorfer, F. Balestro, *Nature*. **453**, 633–637 (2008).
4. C. Brüne et al., *Nat. Phys.* **8**, 485–490 (2012).
5. R. Vincent, S. Klyatskaya, M. Ruben, W. Wernsdorfer, F. Balestro, *Nature*. **488**, 357–360 (2012).
6. M. Weisheit et al., *Science*. **315**, 349–351 (2007).
7. D. Chiba et al., *Phys. Rev. Lett.* **104**, 1–4 (2010).
8. S. O. Valenzuela, M. Tinkham, *Nature*. **442**, 176–179 (2006).
9. I. M. Miron et al., *Nat. Mater.* **9**, 230–234 (2010).
10. R. A. Buhrman, *Science*. **336**, 555 (2012).
11. J. Wang et al., *Science*. **299**, 1719–1722 (2003).
12. N. Fujimura, T. Ishida, T. Yoshimura, T. Ito, *Appl. Phys. Lett.* **69**, 1011–1013 (1996).
13. T. Lottermoser, T. Lonkai, U. Amann, M. Fiebig, *Nature*. **430**, 541–544 (2004).
14. C. A. F. Vaz, *J. Phys. Condens. Matter*. **24** (2012), doi:10.1088/0953-8984/24/33/333201.
15. J. M. Hu, L. Q. Chen, C. W. Nan, *Adv. Mater.* **28**, 15–39 (2016).
16. J. M. D. Coey, *Magnetism and magnetic materials* (Cambridge University Press, 2010).
17. A. Mardana, S. Ducharme, S. Adenwalla, *Nano Lett.* **11**, 3862–3867 (2011).
18. A. Sonntag et al., *Phys. Rev. Lett.* **112**, 017204 (2014).
19. D. Gatteschi, R. Sessoli, J. Villain, *Molecular Nanomagnets* (Oxford University Press, ed. 2, 2006).
20. I. G. Rau et al., *Science*. **344**, 988–92 (2014).
21. C. F. Hirjibehedin et al., *Science*. **317**, 1199–203 (2007).
22. A. S. Zyazin et al., *Nano Lett.* **10**, 3307–3311 (2010).
23. B. W. Heinrich et al., *Nano Lett.* **13**, 4840–4843 (2013).
24. B. W. Heinrich, L. Braun, J. I. Pascual, K. J. Franke, *Nano Lett.* **15**, 4024–4028 (2015).
25. J. J. Parks et al., *Science*. **328**, 1370–1373 (2010).
26. B. Bryant, a Spinelli, J. J. T. Wagenaar, M. Gerrits, a F. Otte, *Phys. Rev. Lett.* **111**, 127203 (2013).
27. J. Martinez-Castro et al., *Nat. Nanotechnol.* **13**, 19–23 (2018).
28. A. F. Otte et al., *Nat. Phys.* **4**, 847–850 (2008).
29. J. C. Oberg et al., *Nat. Nanotechnol.* **9**, 64–8 (2014).
30. J. Repp, G. Meyer, S. Paavilainen, F. Olsson, M. Persson, *Phys. Rev. Lett.* **95**, 225503 (2005).

31. V. Garcia *et al.*, *Nature*. **460**, 81–4 (2009).
32. S. Kitamura, M. Iwatsuki, *Appl. Phys. Lett.* **72**, 3154–3156 (1998).
33. M. R. Kryder *et al.*, *Proc. IEEE*. **96**, 1810–1835 (2008).
34. L. Zhang, T. Miyamachi, T. Tomanić, R. Dehm, W. Wulfhekel, *Rev. Sci. Instrum.* **82**, 103702 (2011).
35. F. J. Giessibl, *Rev. Mod. Phys.* **75** (2003).
36. F. M. Leibsle, S. S. Dhesi, S. D. Barrett, a. W. Robinson, *Surf. Sci.* **317**, 309–320 (1994).
37. I. Horcas *et al.*, *Rev. Sci. Instrum.* **78**, 013705 (2007).

**Acknowledgements**

We thank Markus Ternes for stimulating discussions and Marten Piantek for experimental contributions. J.M.C. and C.F.H. acknowledge financial support from Specs GmbH, EPSRC [EP/H002367/1 and EP/M009564/1], and the Leverhulme Trust [RPG- 2012-754]; D.S. acknowledges funding from MINECO [MAT2016-78293-C6-6-R], FEDER funds through the Interreg-POCTEFA program [TNSI/EFA194/16/], and the use of SAI-Universidad de Zaragoza.

**Author contributions:** J.M.C., D.S., and C.F.H. conceived of the project; J.M.C. and D.S. performed the experiments and analyzed the results; all authors discussed the results and contributed to the writing of the paper.

**Figures and Tables**

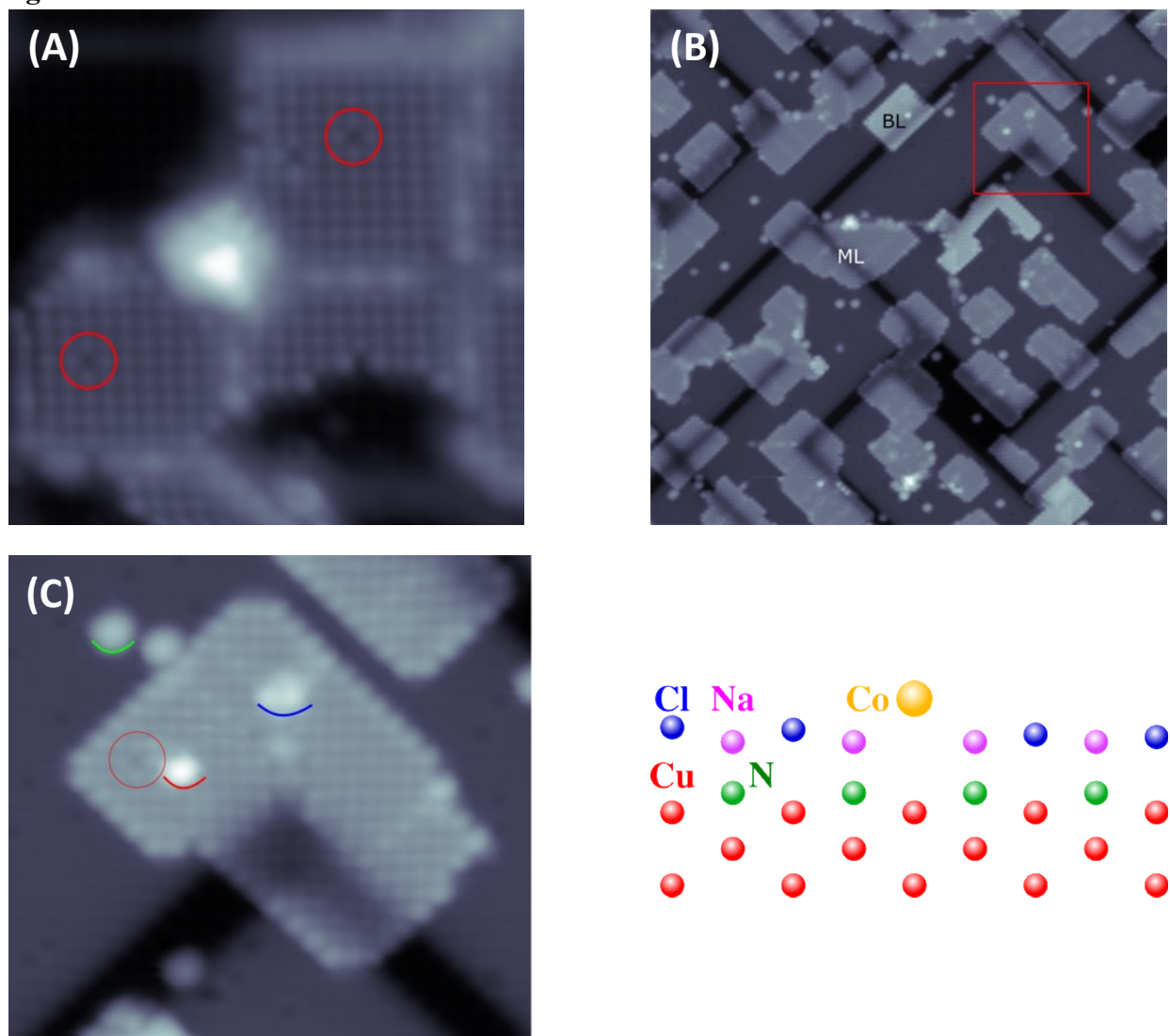

**Fig. 1. Vacancies and Co atoms on NaCl ML.** (**A**) Topographic STM image of a Cu(001) surface mostly covered by Cu$_2$N islands with small stripes of Cu in between and NaCl ML on top (9 nm × 9 nm; $V_{set}$=-1.3 V, $I_{set}$=50 pA). Red circles denote two of the Cl vacancies in the NaCl ML. (**B**) Topographic STM image overview of a Cu(001) surface fully covered by Cu$_2$N with NaCl ML and BL on top with adsorbed single Co atoms (50 nm × 50 nm, $V_{set}$=-1.3 V , $I_{set}$=10 pA). (**C**) Co atoms adsorbed on top of Cu$_2$N (green arc), Na site of NaCl ML (red arc), and a Cl vacancy (blue arc). A nearby Cl vacancy (red circle) is also highlighted (11 nm × 11 nm). (**D**) Side view of an illustration of an NaCl ML on top of Cu$_2$N with Cl vacancy as well as a Co atom on top.

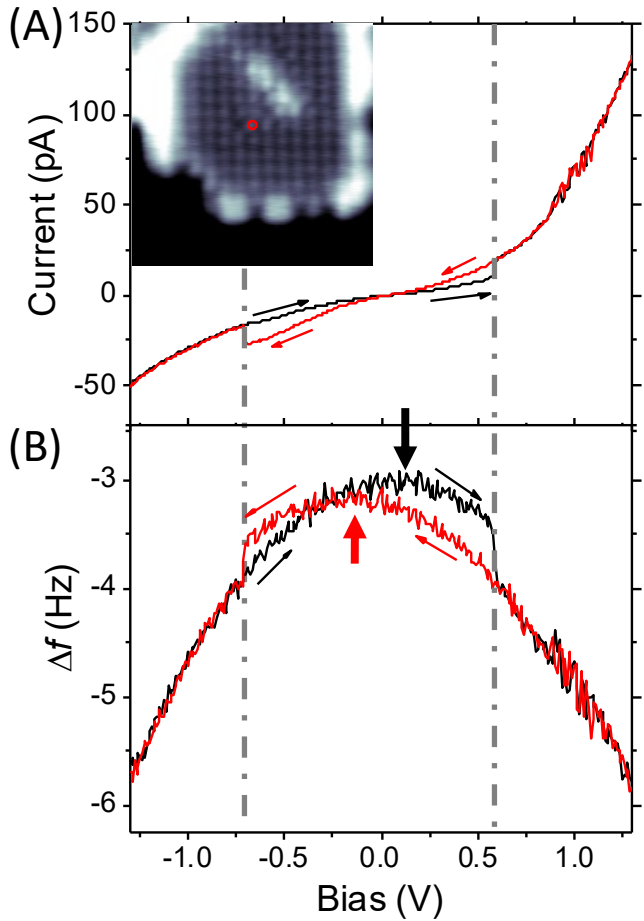

**Fig. 2. Hysteretic reversal of the polarization near a Cl vacancy in the NaCl ML.** **(A)** $I(V)$ acquired above a Cl$^-$ anion near a Cl vacancy in the NaCl ML ($V_{set}$=-1.3 V, $I_{set}$=50 pA). Two states can be distinguished: a self-poled initial state (black) and a switched state (red). The sharp change in current indicates the switching of the polarization (dashed grey vertical lines). Inset: topographic STM image of the NaCl ML with the Cl vacancy. Red circle denotes the position where the spectroscopy was performed (5.8 nm × 5.8 nm, $V_{set}$=-1.3 V, $I_{set}$=50 pA). **(B)** Simultaneously acquired $\Delta f(V)$. The change in the polarization of the NaCl ML is observed in a clear shift of $V_{cpd}$ between the two states as indicated by the vertical black and red arrows, which mark the maximum of the respective parabolas.

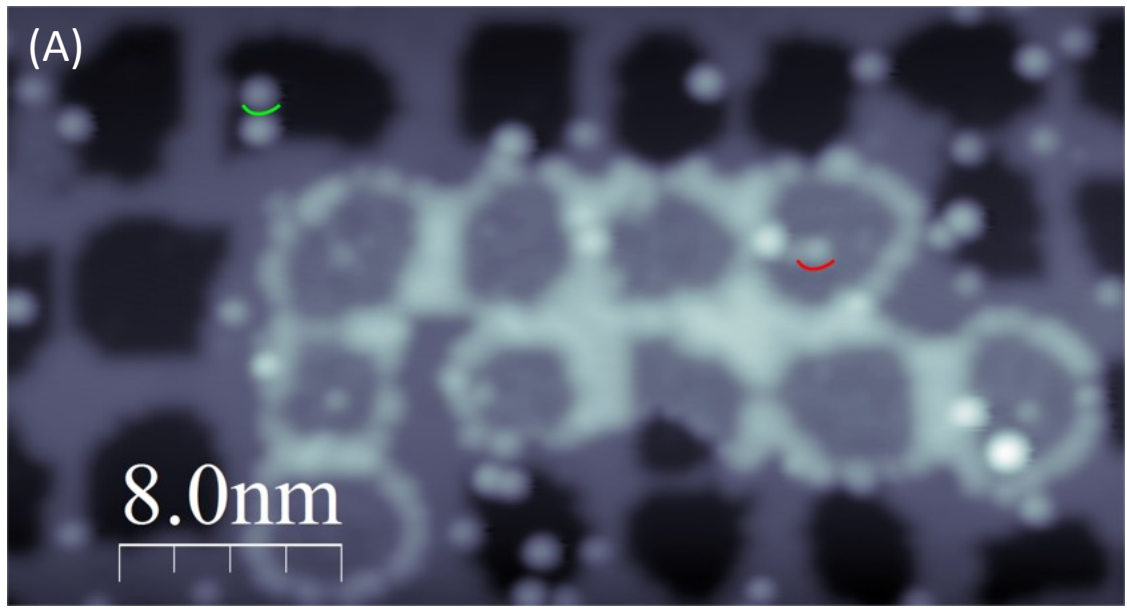

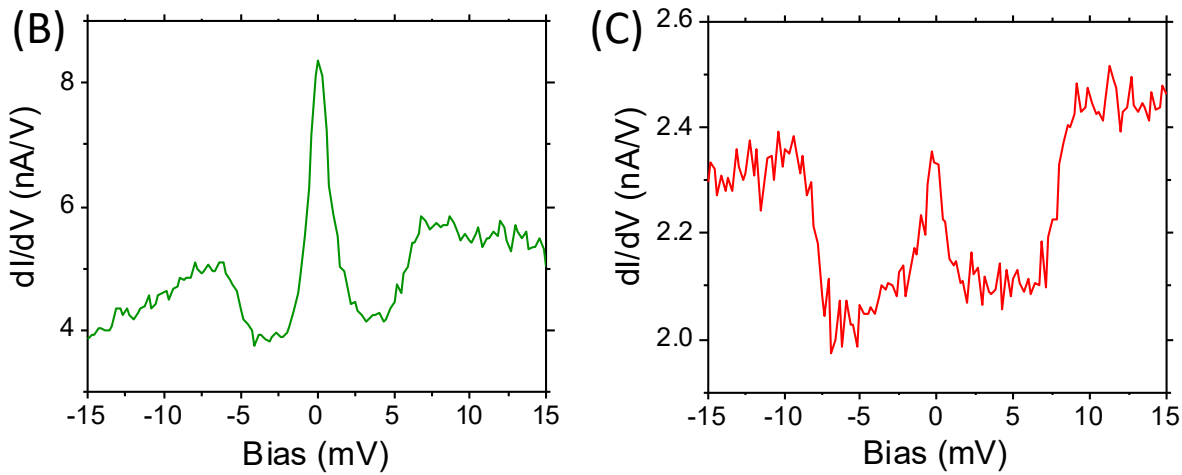

**Fig. 3. Inelastic electron tunneling spectroscopy for Co atoms on Cu$_2$N and NaCl ML.** **(A)** STM topographic image of a Cu$_2$N/Cu(001) surface and a partial coverage of NaCl ML. Co atoms can be observed on top of both the Cu$_2$N (green arc) and NaCl ML (red arc). (40 nm × 20 nm, $V_{set}$=-1.3 V, $I_{set}$=10 pA). **(B)** Typical low energy spectroscopy of a Co atom on top of Cu$_2$N (*28*). A characteristic Kondo resonance peak centered at $V$=0 is observed along with conductance steps due to inelastic spin-flip excitations ($V_{set}$=-15 mV, $I_{set}$=80 pA, $V_m$=150 µV). **(C)** Low energy spectroscopy of a Co atom on NaCl ML close to a Cl vacancy obtained with the same time that was used to measure the spectrum in panel B. As for Co on Cu$_2$N, a Kondo resonance and inelastic tunneling steps can be observed ($V_{set}$=-15 mV, $I_{set}$=30 pA, $V_m$= 150 µV).

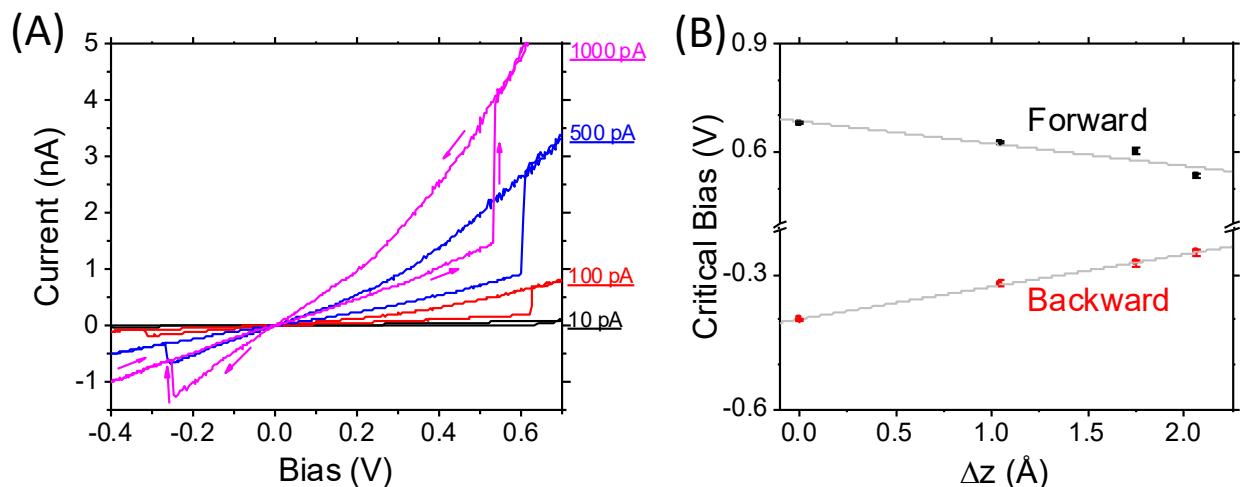

**Fig. 4. Electric field induced polarization switching above a Co atom adsorbed near a Cl vacancy. (A)** $I(V)$ spectroscopy of the polarization switching cycle for a Co atom near a Cl vacancy in the NaCl ML performed for different values of $I_{set}$ to tune the change in tip-sample distance $\Delta z$. The relative height variation was extracted from the change in the relative piezo movement. A critical bias is observed when the current step occurs, marking the transitions between the two polarization states. **(B)** Change in critical bias vs $\Delta z$ with best-fit lines corresponding to critical electric fields $E_c$=0.6 GV m$^{-1}$ for positive applied electric fields and $E_c$=0.7 GV m$^{-1}$ for negative applied electric fields. Vertical error bars represent the width of the step in $I(V)$.

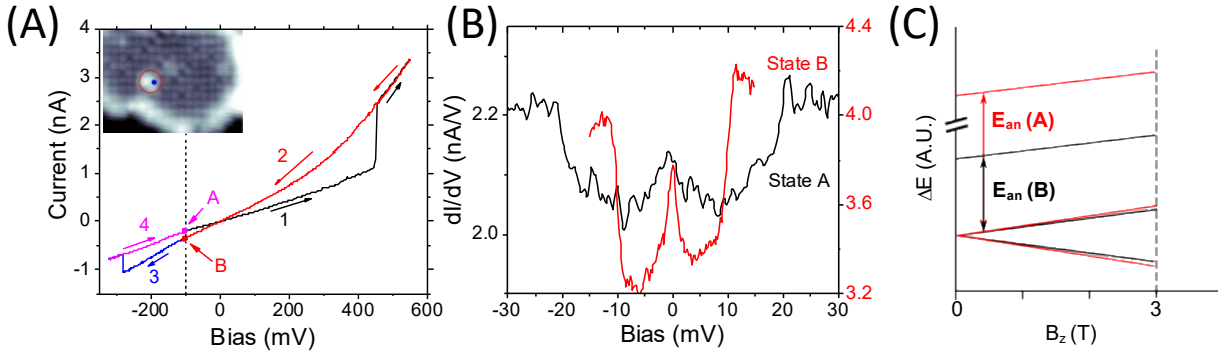

**Fig. 5. Controlling MAE with polarization switching. (A)** Hysteretic polarization switching cycle performed over an adsorbed Co atom near a Cl vacancy in the NaCl ML. Spectroscopy begins at point A. As the voltage is increased along line 1, a polarization switch is induced. The voltage is the decreased along line 2 back to the same voltage as point A, but now the current is different (point B). The voltage is then further decreased along line 3 until the polarization switches back to the original state. Finally, as the voltage is increased along line 4 the spectrum returns to point A. Points A and B also indicate the position in the cycle where IETS is performed. Inset: topographic STM image of the measured Co atom on the NaCl ML system (7 nm × 5 nm, $V_{set}$=-1.3 V, $I_{set}$=10 pA). **(B)** Low-bias d$I$/d$V$ spectroscopy performed for the Co atom near a Cl vacancy in the NaCl ML in the two different polarization states. The MAE, as measured from the voltage of the symmetric steps in d$I$/d$V$, is significantly different in each case. ($I_{set}$=200 pA for state A and 370 pA for state B, $V_{set}$= -100 mV). **(C)** Energy level diagram for an $S$=3/2 system in the presence of negative axial magnetic anisotropy as a function of out-of-plane magnetic field for two different MAE values, corresponding to the substrate polarization in state A (black) and B (red).

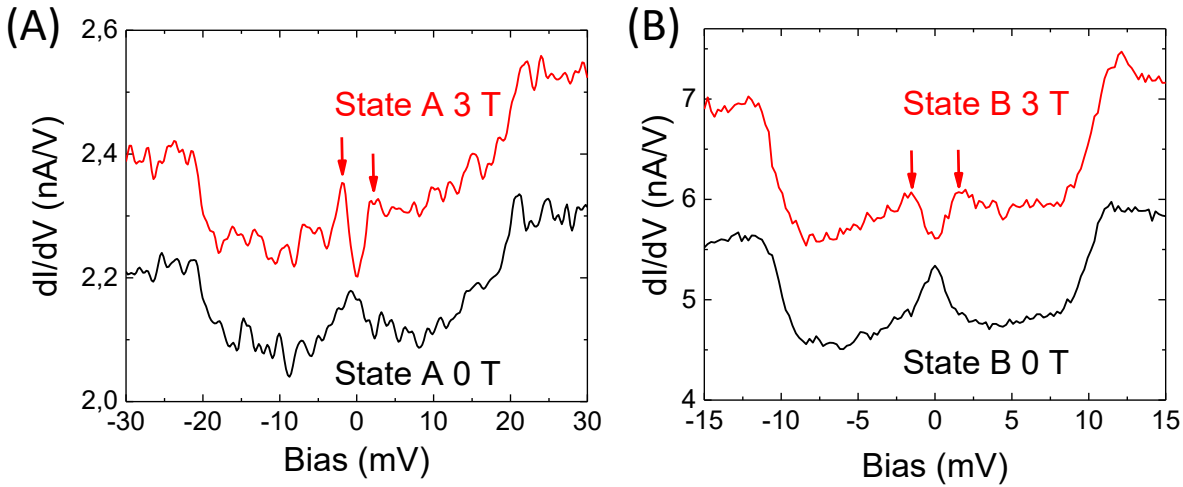

**Fig. 6. Magnetic field dependence of spin-excitation energy for the self-poled and switched polarization state. (A)** Low energy spectroscopy of a Co atom adsorbed on NaCl ML near a Cl vacancy measured at $B=0$ and $B=3$ T (shifted vertically for clarity) with the NaCl ML in the self-poled state (state A). A distinct Zeeman splitting of the Kondo resonance (vertical red arrows) near $E_F$ is observed at 3 T, as has been observed for Co on bare Cu$_2$N (28). A slight shift of the inelastic steps with magnetic field, which is expected to be very small for these relatively small changes in $B$ (28), can also be discerned. ($V_{set}$=-100 mV, $I_{set}$=53 pA, $V_m$=125 µV) **(B)** Same as panel A but for the switched state (state B). ($V_{set}$=-100 mV, $I_{set}$=61 pA, $V_m$=125 µV).

**Supplementary Materials**

Figures S1-S2

**Supplementary Materials**

**Figures S1-S2**

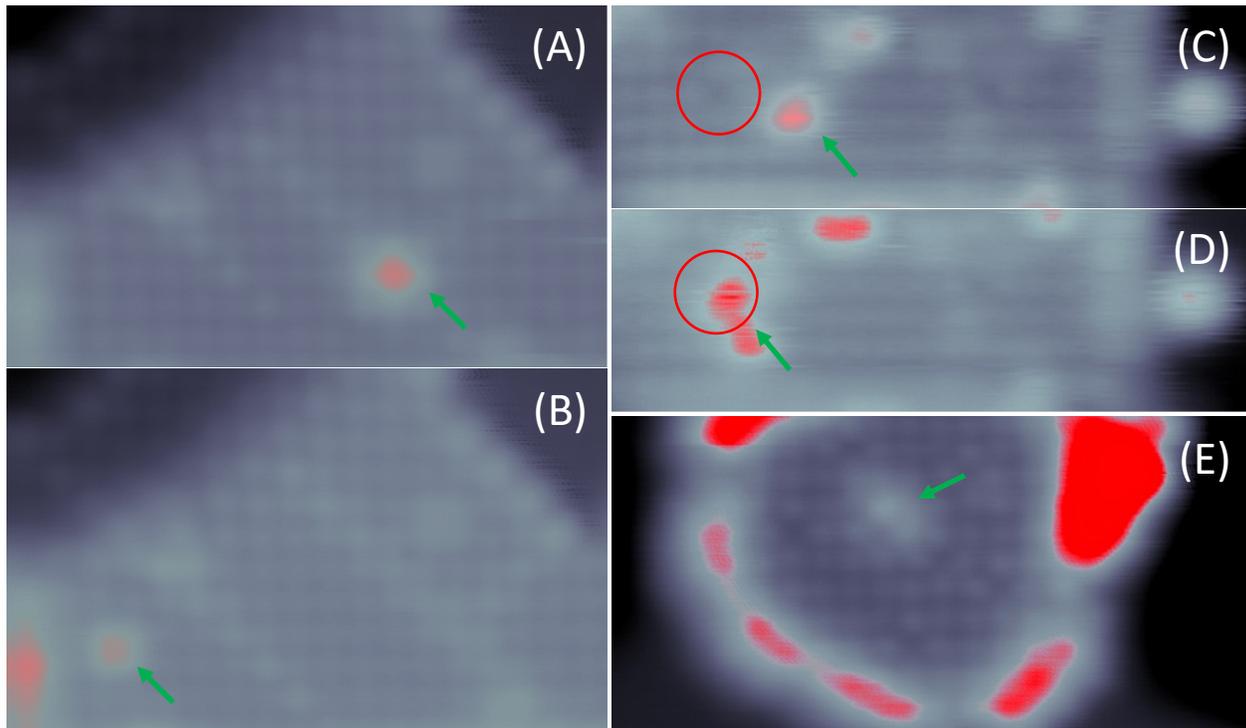

**Fig. S1.** Adsorption and manipulation of Co atoms on NaCl ML/Cu$_2$N. **(A)** Co atom brought to a Cl site with lateral manipulation (marked with the green arrow). (7 nm × 4 nm). **(B)** Co atom from panel A brought to a Na site (marked with the green arrow) (7 nm × 4 nm). **(C)** Co atom sitting on top of a Na site. A Cl vacancy is highlighted with a red circle while the Co atom is marked with the green arrow (10 nm × 2.2 nm). **(D)** Co atom brought close to a Cl vacancy, resulting in a change in its apparent shape and size. (10 nm × 2.2 nm). **(E)** Co atom naturally adsorbed on top of a Cl vacancy. The adsorbed Co atom can be identified as the cloud-like feature (green arrow) while the Cl vacancy has the characteristic four lobe structure (10 nm × 4 nm). All topographies were measured at $V_{set}$=-1.3 V, $I_{set}$=10 pA except panel D with $I_{set}$=50 pA. All manipulations were initialized at $V_{set}$= -30mV, $I_{set}$=10 pA.

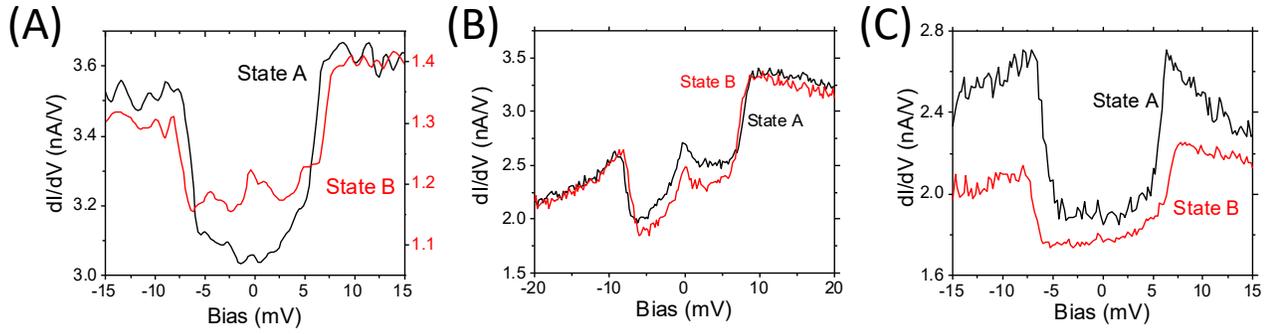

**Fig. S2.** Low energy spectroscopy of different on Co atoms adsorbed near a Cl vacancy in the NaCl ML. **(A-C)** Low-bias d$I$/d$V$ spectroscopy performed for three different Co atoms near a Cl vacancy in the NaCl ML in the two different NaCl polarization states. The general features are the same in all cases, though the MAE either increases or decreases from state A to B depending on the details of the local environment, which cannot be resolved by STM imaging. We note that the main factor determining the single-ion MAE is the charge distribution around an atom, which fixes the crystal electric field and influences the orbital configuration. Similar variations are also observed for changes in the Kondo resonance. The setpoint conditions are (A) $V_{set}$=-20 mV, $I_{set}$=50 pA (State A) and $I_{set}$=20 pA (State B), $V_m$=125 µV; (B) $V_{set}$=-20 mV, $I_{set}$=50 pA, $V_m$=125 µV; (C) $V_{set}$=-15 mV, $I_{set}$=30 pA, $V_m$=125 µV.